# Local magnetic anisotropy controlled by a surface nano-modulation

J. Briones, F. Montaigne, G. Lengaigne, D. Lacour[a)], M. Hehn

*Institut Jean Lamour, CNRS - Nancy-Université - UPV-Metz, boulevard des aiguillettes*
*BP 70239, F-54506 Vandoeuvre lès Nancy, France*

A topological modulation of magnetic thin films can induce a magnetic anisotropy of magnetostatic origin. In this letter, we report on the magnetic properties of NiFe layers deposited on wavy shaped Si substrates. Without any modulation, our films always present an intrinsic anisotropy. We show unambiguously that patterning the substrate can overcome this anisotropy and even impose a different easy axis of magnetization. This allows the definition of two orthogonal easy axes at different places on the same substrate. This control of anisotropy both in direction and intensity paves the way to the realization of high precision bidimensional magnetic sensors.
*Pacs numbers: 75.30.-m ; 75.60.Jk ; 75.75.+a ; 75.47.-m ; 85.75.Ss ; 85.75.-d*

The behaviour of many spin electronics devices results from the influence of the magnetic configuration on the electrical conduction. The magnetic configuration itself is a complex function of external parameters (magnetic field, temperature, electrical currents…) and material parameters (magnetization, exchange constant, anisotropy…). Among these last parameters, magnetic anisotropy is certainly the most important since it is the main mean to control the electrical response to a given external field. Well defined uniaxial anisotropies are essential for good definition and retention of digital information in magnetic recording or magnetic memories. It is also necessary for magnetic field sensors to obtain a linear hysteresis-free response [1]. In this last case, the anisotropy defines both the sensing direction and the sensitivity.

Fortunately, magnetic anisotropy is also the most variable parameter for a given material and small structural variations can induce important magnetic anisotropy. Consequently various ways have been used to introduce a magnetic anisotropy [2]. An uniaxial anisotropy can be induced by the geometry of the growth system, the application of a magnetic field during or after the growth [3], the coupling with an other anisotropic layer [4] or the use of an anisotropic substrate [5]. The amplitude and even the direction of the anisotropy results from a complex interplay between volume magneto-crystalline anisotropy reflecting the symmetry of the crystal structure [2], anisotropies linked to the break of symmetry at the magnetic layers interfaces or defects [6], magneto-elastic anisotropy linked to the presence of stress [7] and many other aspects. Therefore the exact value of the anisotropy is generally poorly tuneable. Furthermore these anisotropies are defined at the "wafer level" and therefore simultaneous fabrication of devices having different anisotropies is impossible. Use of lithography techniques circumvent this difficulty by defining locally structures of different orientation or size. Lithographically defined large aspect ratio structures exhibit adjustable anisotropies originating mainly from magnetostatic contribution [8]. Keeping continuous thin film geometry, Oepts et al. grown NiFe films on pre-structured substrates [9]. The InP substrates exhibit a modulated topology (called V-grooves) defined by Ebeam lithography and anisotropic chemical etching. The origin of the high anisotropy fields (above 300 Oe) remains unclear and is not purely magnetostatic.

In this letter we study the magnetic anisotropy induced by a "wavy" silicon surface in a NiFe film. Modulations of surface with 300 and 400 nm periods and less than 66 nm deep induce a magnetic anisotropy in the 26 - 62 Oe range. The interplay of this "controlled" anisotropy with the "intrinsic" anisotropy due to the growth geometry is studied. Local tuning of the anisotropy field in amplitude and direction is demonstrated and the anisotropy field is found to vary linearly with modulation depth.

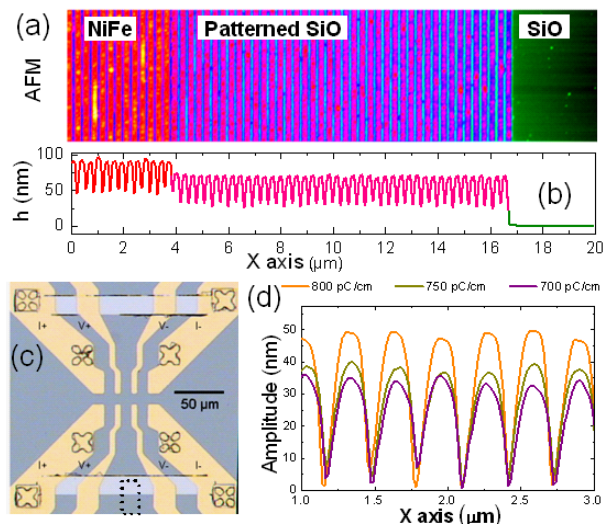

FIG. 1. (color on line) (a) AFM image recorded on a completed device. (b) The AFM profile shows the conformal deposition of the magnetic layer. The period of modulation is the same 306 nm and the amplitudes differ by less than 10%. (c) Optical photography of the completed devices. The black dots surround the AFM measurements zone. (d) Cross sections along the direction of modulation of Si surfaces processed using different exposition doses.

During a first step, a Si surface is patterned by $Ar^+$ ion milling. Periodic arrays of lines are defined by Ebeam lithography (JEOL 6500F converted SEM) in a MMA/PMMA bilayer resist. Due to the high density of the pattern, the exposure dose is limited by proximity effects and overdeveloping of the MMA layer. For a 300 nm

[a)] Electronic mail: lacour@lpm.u-nancy.fr

period the linear dose typically ranges from 700 to 800 pC/cm for a 30 keV electron energy. An Al layer is then thermally evaporated and subsequently lifted-off in acetone. The Si surface is etched by a 200 eV Ar$^+$ ion beam at 45° incidence. The etching time is long enough so that the aluminium film is completely removed and that the etched thickness is not determined by the etching time but by the Al mask thickness. Lines are defined in two orthogonal directions using several different doses on the same sample. Therefore the effect of different modulations directions on the same magnetic thin film can be studied. In a second step, two 150 µm x 20 µm wide windows are opened by E-beam lithography in a MMA/PMMA bilayer at locations where the Si substrate has been patterned. The multilayer stack Ta(5 nm)/Ni$_{81}$Fe$_{19}$(10 nm)/Pt(3 nm) is then deposited by sputtering. After the resist lift off, the magnetic layer subsists only on the modulated substrate regions. During the last step, macroscopic Ti/Au leads for magneto-transport measurements are added using a UV lithography process.

Figure 1a represents the AFM characterisation obtained on a completed device. The profile plotted in Fig. 1b shows the conformal deposition of our multilayer on a patterned Si substrate. It can be remarked that large unprotected areas *i.e.* without stripes (right part of Fig. 1 a and b) are deeper etched than the others. This difference is attributed to a redeposition effect lowering the effective etching rate in the case of dense patterns. Profiles obtained for different doses are represented in Fig 1d. The amplitude of modulation increases with the dose. This variation is explained by the reduced Al deposition rate in the smaller (lower dose) structures due to a non exactly normal angle of deposition.

DC resistance measurements as a function of applied field have been performed using a conventional 4 probes measurement with an applied DC current of 700 µA. In the so called 'transverse' 150 µm x 20 µm area, the lines are orthogonal to the current direction while in the so called 'longitudinal' area, the lines are parallel to the current direction. Those geometries will be referenced as the transverse and longitudinal geometries respectively. In a same way, when a in plane magnetic field is applied perpendicularly (respectively along) to the current direction, the field is labelled transverse (respectively longitudinal).

We have focused our study on the magnetic properties of Ta(5)/Ni$_{81}$Fe$_{19}$(10)/Pt(3) multilayers taking profit from the magnetoresistive properties of the Ni$_{81}$Fe$_{19}$. Indeed, this stack exhibits anisotropy of magneto-resistance (AMR) equal to 0.65%. AMR measurements done with the magnetic field applied perpendicularly to the easy direction provides a quantitative determination of the anisotropy field. Considering a single domain behaviour of the Ni$_{81}$Fe$_{19}$ layer, the resistivity, $r$, can be expressed as [10]:

(i)
$$r(H_{transverse}) = r_{//} + (r_{\perp} - r_{//})\left(\frac{H}{H_a}\right)^2$$
$$r(H_{longitudinal}) = r_{//}$$

when the anisotropy field $H_a$ is parallel to the current and is equal to:

(ii)
$$r(H_{transverse}) = r_{\perp}$$
$$r(H_{longitudinal}) = r_{\perp} - (r_{\perp} - r_{//})\left(\frac{H}{H_a}\right)^2$$

when the anisotropy field $H_a$ is perpendicular to the current.

$r_{//}$ and $r_{\perp}$ are the layer resistivities when the magnetization is parallel or perpendicular to the current; $r(H_{transverse})$ and $r(H_{longitudinal})$ are respectively the layers resistivities for two different directions of applied magnetic field: the magnetic field is either transverse (noted $H_{transverse}$) or longitudinal ($H_{longitundinal}$) to the long axis of the NiFe stripes.

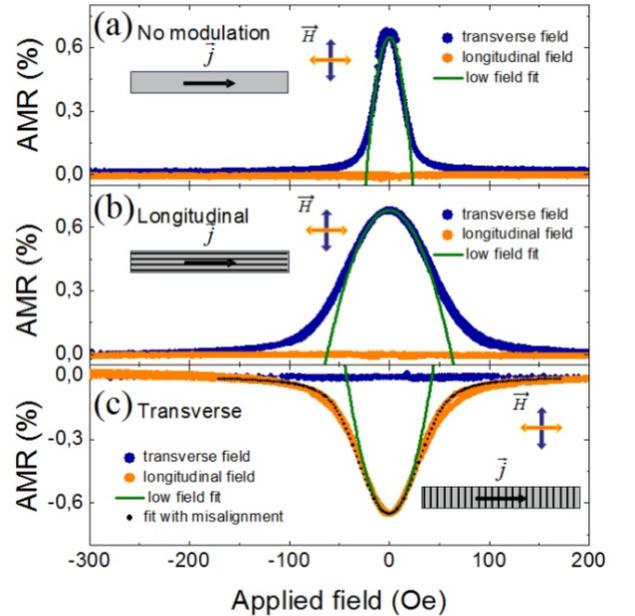

FIG. 2. (color on line) Variation of AMR with in plane field applied either along the current direction (labeled longitudinal field) either perpendicularly to the current direction (labeled transverse field). (a) Without any topological modulation. (b) In the longitudinal geometry (steps parallel to the current direction). (c) In the transverse geometry (steps perpendicular to the current direction). The schematic reminds to the reader the device configuration, the field and current directions as well as the array of lines transferred to the Si substrate.

Without any modulation, our films present an intrinsic anisotropy field as shown in the figure 2a. Considering both the measurements with transverse and longitudinal fields, it appears that case (i) applies. Consequently the direction of anisotropy is parallel to the current direction. The source of this intrinsic anisotropy originates either from the film deposition in the target magnetron magnetic field, either from a columnar growth of the film during substrate motion over the Ni$_{81}$Fe$_{19}$ target used to promote the film thickness uniformity. Fitting the $r(H_{transverse})$ curve for field values less than 20 Oe allows to extract an intrinsic anisotropy field of 22 Oe (— on Fig. 2a).

Patterning the Si substrate with a period of 300 nm and an amplitude equal to 34 nm in the longitudinal geometry reinforces the anisotropy field as exemplified in figure 2b. First of all, considering the measurements with transverse and longitudinal fields, it appears that the anisotropy direction is still along the current direction. The fit of the $r(H_{transverse})$ curve for field values around zero allows to extract an anisotropy field of 62 Oe ( — on Fig. 2b ).

Patterning the Si substrate with a period of 300 nm and an amplitude equal to 39 nm in the transverse geometry has a strong impact on the anisotropy field as exemplified in figure 3c. Considering the measurements with transverse and longitudinal fields, it appears that case (i) does not apply anymore and case (ii) has to be considered: $H_a$ is perpendicular to the current direction! Then, fitting the $r(H_{longitudinal})$ curve for field values around zero allows to extract an anisotropy field of 42 Oe ( — on Fig. 2c ).

From this set of measurements, we can conclude that the easy axes of magnetization are along the steps directions in both cases. So, we show unambiguously that patterning the substrate can counter the intrinsic anisotropy and even impose a different easy axis of magnetization. Note that, for magnetic field values close to the anisotropy fields, the resistivity is not quadratic-in-field anymore. This deviation is commonly observed and attributed to a misalignment between the applied field and the local easy direction of magnetization [10]. For instance Fig. 2c presents a fit of an AMR curve taking into account an effective misalignment of 10°.

For thinner NiFe films the effect of the surface nano-modulation is drastically decreased. The intrinsic anisotropy is not overcame by the surface modulation in the transverse geometry. A batch of AMR measurements done on 5 and 3 nm thick NiFe films (not presented in this letter) show the appearance of magnetic ripples in the range of studied field wixh could be responsible.

Our last point concerns the control of the anisotropy field intensity. Keeping the NiFe film thickness equal to 10 nm, we have varied the amplitude and period of the Si substrate modulation as shown in figure 3. In each case, the anisotropy field shows a linear variation with the modulation amplitude. Furthermore, when the modulation period increases (-⊙- on Fig. 3), the slope decreases. As a matter of fact, the anisotropy field intensity can be tuned with simple rules. As expected in the longitudinal geometry (-●- and -⊙- on Fig. 3), a decrease of the modulation amplitude leads to convergence of Ha values toward 22 Oe (the intrinsic anisotropy field) and this independently on the modulation period. For a topological modulation perpendicular to the intrinsic anisotropy field ( -▲- on Fig. 3), the competition between the intrinsic anisotropy and the anisotropy induced by a nano-modulation of the surface gives rise to a more complex behaviour [11].

In this study, we have unambiguously shown that the easy axis of magnetization of a 10 nm thick NiFe layer can be locally controlled by a surface nano-modulation of a Si substrate. The easy axis directions are parallel to the step directions and the anisotropy field can be brought to be equal in each chosen direction. The intensity of the anisotropy field can be tuned by simple rules : changing either the line array periodicity or the depth of Si etching. The control on the same wafer of the anisotropy field of two areas separated by only several microns supporting the same magnetic thin film is one of the keys for the design of high precision bidimensional magnetic sensors.

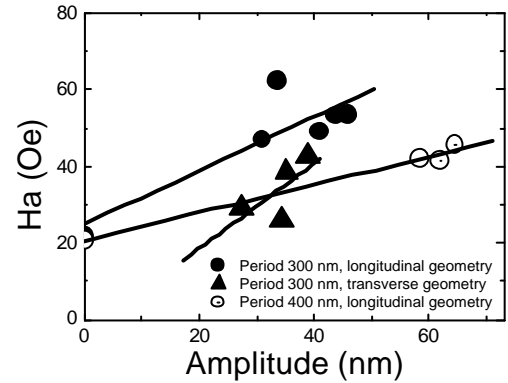

FIG. 3. Variation of the anisotropy field as a function of the modulations amplitude measured by AFM. The lines are guides for the eyes.

The work presented in this letter is partly supported by La Région Lorraine. J.B. acknowledges support from CONACYT. The authors also acknowledge M. Alnot and S. Girod.